\documentclass[%
 aip,
 graphicx,
 amsmath,amssymb,
 reprint,%
]{revtex4-1}

\usepackage{graphicx}
\usepackage{dcolumn}
\usepackage{bm}

\usepackage[utf8]{inputenc}
\usepackage[T1]{fontenc}
\usepackage{mathptmx}
\usepackage{etoolbox}
\usepackage{color}
\usepackage{multirow}
\usepackage{ulem}
\usepackage{comment}
\usepackage{dcolumn}
\usepackage{siunitx}

\usepackage[acronym, shortcuts]{glossaries}
\newacronym{CPT}{CPT}{coherent population trapping}
\newacronym{MEMS}{MEMS}{micro-electro-mechanical systems}
\newacronym{ASG}{ASG}{aluminosilicate glass}
\newacronym{BSG}{BSG}{borosilicate glass}
\newacronym{ALD}{ALD}{atomic layer deposition}

\makeatletter
\def\@email#1#2{%
 \endgroup
 \patchcmd{\titleblock@produce}
  {\frontmatter@RRAPformat}  {\frontmatter@RRAPformat{\produce@RRAP{*#1\href{mailto:#2}{#2}}}\frontmatter@RRAPformat}
  {}{}
}%
\makeatother
\begin{document}

\preprint{AIP/123-QED}

\title{On the reduction of gas permeation through the glass windows of micromachined vapor cells using Al$_2$O$_3$ coatings}

\author{C. Carl\'e}
\affiliation{FEMTO-ST Institute, CNRS, Universit\'e de Franche-Comt\'e, Besançon, France}
\author{A. Mursa}
\affiliation{FEMTO-ST Institute, CNRS, Universit\'e de Franche-Comt\'e, Besançon, France}
\author{P. Karvinen}
\affiliation{University of Eastern Finland, Department of Physics and Mathematics, Joensuu, Finland}
\author{S. Keshavarzi}
\affiliation{FEMTO-ST Institute, CNRS, Universit\'e de Franche-Comt\'e, Besançon, France}
\author{M. Abdel Hafiz}
\affiliation{FEMTO-ST Institute, CNRS, Universit\'e de Franche-Comt\'e, Besançon, France}
\author{V. Maurice}
\affiliation{IEMN, Universit\'e Lille, CNRS, Centrale Lille, Universit\'e Polytechnique Hauts-de-France, Lille, France}
\author{R.~Boudot}
\affiliation{FEMTO-ST Institute, CNRS, Universit\'e de Franche-Comt\'e, Besançon, France}
\author{N. Passilly}
\affiliation{FEMTO-ST Institute, CNRS, Universit\'e de Franche-Comt\'e, Besançon, France}
\email{nicolas.passilly@femto-st.fr}

\date{\today}

\begin{abstract}

Stability and precision of atomic devices are closely tied to the quality and stability of the internal atmosphere of the atomic vapor cells on which they rely. Such atmosphere can be stabilized by building the cell with low permeation materials such as sapphire, or aluminosilicate glass in microfabricated devices. Recently, we showed that permeation barriers made of Al$_2$O$_3$ thin-film coatings deposited on standard borosilicate glass could be an alternative for buffer gas pressure stabilization. In this study, we hence investigate how helium permeation is influenced by the thickness, ranging from \SI{5} to  \SI{40}{\nano\meter}, of such Al$_2$O$_3$ thin-films coated by atomic layer deposition. Permeation rates are derived from long-term measurements of the pressure-shifted transition frequency of a \ac{CPT} atomic clock. From thicknesses of \SI{20}{\nano\meter} onward, a significant enhancement of the cell hermeticity is experienced, corresponding to two orders of magnitude lower helium permeation rate. In addition, we test cesium vapor cells filled with neon as a buffer gas and whose windows are coated with  \SI{20}{\nano\meter} of Al$_2$O$_3$. As for helium, the permeation rate of neon is significantly reduced thanks to alumina coatings, leading to a fractional frequency stability of \SI{4}~$\times$~10$^{-12}$ at 1 day when the cell is used in a \ac{CPT} clock. These features outperform the typical performances of uncoated Cs-Ne borosilicate cells and highlight the significance of Al$_2$O$_3$ coatings for buffer gas pressure stabilization.

\end{abstract}
\maketitle

\section{\label{sec:level1}Introduction}
The convergence of \ac{MEMS}, integrated photonics and atomic spectroscopy has  led to the development of chip-scale atomic devices~\cite{Kitching:APR:2018}  that are both wafer-scalable and of high performance such as microwave atomic clocks~\cite{Knappe:APL:2004, Knappe:Elsevier:2007, Yanagimachi:APL:2020, Batori:PRap:2023, Carle:OE:2023, Martinez:NC:2023}, optical frequency references~\cite{Newman:O:2019, Maurice:OE:2020, Newman:OL:2021, Gusching:OL:2023, Dyer:PRAp:2023}, magnetometers~\cite{Shah:Nature:2007, Griffith:OE:2010, Boto:Nature:2018}, voltage references~\cite{Teale:AVS:2022}, or optical isolators~\cite{Levy:2020}. It is also responsible for the emergence of atomic diffractive optical elements~\cite{Stern:NC:2019}, quantum memories~\cite{Treutlein:2023} or  chip-scale laser-cooling platforms~\cite{McGilligan:APL:2020, McGilligan:RSI:2022}.

The core of such devices consists usually of a microfabricated alkali vapor cell, designed to provide a stable atmosphere of alkali vapor, sometimes combined with a buffer gas pressure. In addition to being compatible with collective fabrication, the cell should ensure both chemical neutrality with respect to the reactive alkali metal, and sufficient hermeticity against leakage or insertion of undesired gas. 

A \ac{MEMS} vapor cell is typically made of cavities etched in silicon by dry or wet etching, which are sandwiched between two anodically-bonded glass substrates~\cite{Kitching:APL:2002}. Based on demonstrated long-term performances, this wafer-stack structure has been now widely adopted, with variants for forming and filling/sealing the cavities~\cite{Bopp:JPP:2021, Dyer:JAP:2022, Maurice:NMN:2022, Lucivero:OE:2022}. 

Especially, various approaches for filling the cell with alkali vapor have been proposed. This includes, e.g., direct pipetting of pure alkali in a preform~\cite{Liew:APL:2004}, chemical reaction between an alkali compound and a reduced agent~\cite{Knappe:OL:2005, Bopp:JPP:2021}, alkali dissociation of a pre-embedded pill or paste dispenser using laser or resistive heating~\cite{Douahi:EL:2007, Maurice:APL:2017, Vicarini:SA:2018}, or dissociation of deposited alkali azide with ultraviolet-light~\cite{Liew:APL:2007, Woetzel:2013, Karlen:OE:2017}.

Regardless of the method adopted to develop the \ac{MEMS} cell, once it is sealed, its inner atmosphere can evolve, inducing some instabilities of the atomic clock or sensor. For instance, one of the involved processes is gas permeation through the cell glass windows~\cite{Norton:1957, Altemose:1961, Rushton:2014}. In Ref.~\cite{Abdullah:APL:2015}, Ne permeation through \ac{BSG} windows of a microfabricated Cs cell, heated at 81$^{\circ}$C, was found to limit the fractional frequency stability of a clock at the level of 5~$\times$~10$^{-11}$ at 1~day integration time, corresponding to a Ne pressure change of about $-$~0.9~$\mu$bar/day. Helium gas is even more critical, given its low density and its non-negligible natural concentration in the Earth's atmosphere. The impact of helium permeation onto the long-term stability of microwave~\cite{Camparo:TIM:2005} or optical~\cite{Lemke:MDPI:2022} Rb cell frequency standards was also emphasized. 

Lately, use of \ac{ASG} has been identified to reduce He permeation in microfabricated vapor cells by more than two orders of magnitude, in comparison with \ac{BSG}~\cite{Dellis:OL:2016, Carle:JAP:2023}. \ac{ASG} was also employed to reduce the contribution of Ne permeation on the frequency stability of a Cs microcell \ac{CPT}-based clock at a level below 2~$\times$~10$^{-12}$ at 1~day~\cite{Carle:OE:2023}. 

Nevertheless, the unique commercial provider of \ac{ASG} wafers, along with their higher cost, makes the search for alternative solutions a stimulating objective. 

Formerly, thin films made of Al$_2$O$_3$ have been employed to mitigate the diffusion of alkali into glass, as well as to prevent reactions between alkali and glass components, thereby increasing the lifetime of microfabricated cells~\cite{Woetzel:2013, Karlen:OE:2017, Pate:OL:2023}. 

Recently, we showed that these Al$_2$O$_3$ thin-film coatings can also be an efficient mean for the reduction of permeation of helium gas when it is deposited onto \ac{BSG} as well as on \ac{ASG}~\cite{Carle:JAP:2023}. For instance, the He permeation rate through Al$_2$O$_3$-coated \ac{BSG} was found to be reduced by a factor 130 compared to uncoated \ac{BSG}, i.e., only 4 times less than through uncoated \ac{ASG}. In this previous study, where uncoated \ac{BSG} and \ac{ASG} windows were compared to Al$_2$O$_3$-coated ones, we used systematically \SI{20}{\nano\meter}-thick films. Indeed, this thickness value was shown to be protective enough for the Cs population in Ref.~\cite{Woetzel:2013} whereas still compatible with anodic bonding. Consequently, it has also been employed in subsequent studies involving other alkali metals such as rubidium~\cite{Karlen:OE:2017} or strontium~\cite{,Pate:OL:2023}. In Ref.~\cite{Woetzel:2013}, several thicknesses, i.e., 3, 6, 11 and \SI{22}{\nano\meter} were tested, and it was reported that \SI{6}{\nano\meter} was already a good protection while only a slight improvement with 11 and \SI{22}{\nano\meter}-thick coatings was observed.

Nevertheless, when looking at the gas permeation, the different atomic species involved may react differently. This paper aims consequently at investigating the impact of the thickness of Al$_2$O$_3$ layers, ranging from \SI{5}{\nano\meter} to \SI{40}{\nano\meter}, on helium gas permeation in microfabricated Cs cells built with \ac{BSG} windows. We show that the permeation reduction factor is improved with increased thickness of the Al$_2$O$_3$ layer, but with a strong improvement noticed for thicker layers than in the case of alkali consumption limitation~\cite{Woetzel:2013}.
In addition, in order to confirm further the interest of using Al$_2$O$_3$ layers as a buffer gas permeation barrier, we fabricated  Al$_2$O$_3$-coated \ac{BSG} Cs microcell filled with neon as buffer gas, and demonstrate a \ac{CPT} clock providing a fractional frequency stability at 1~day below 4~$\times$~10$^{-12}$. These performances, that we believe now limited by light-shift effects, far exceed the ones typically obtained ($\sim$~2~$\times$~10$^{-11}$) using a Cs-Ne \ac{BSG} microcell without Al$_2$O$_3$ coatings~\cite{Vicarini:UFFC:2019, MAH:APL:2022}.

\section{\label{sec:thickness}Dependence on the thickness of A\lowercase{l}$_2$O$_3$ layers}

\begin{figure*}[t]
\centering
\includegraphics[width=\linewidth]{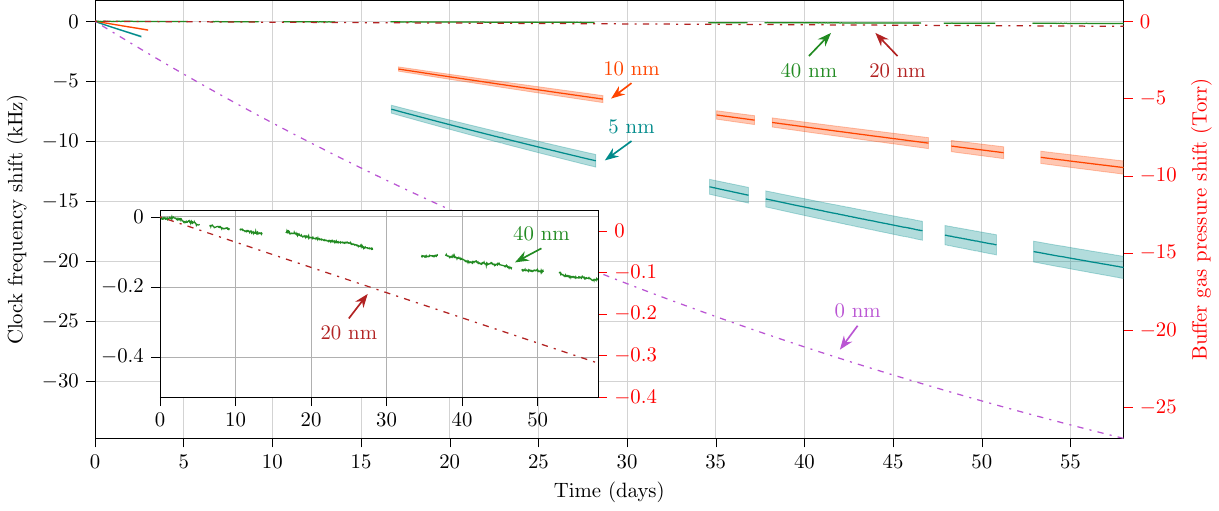}
\caption{Temporal trace of the clock frequency for Cs-He cells made of \ac{BSG} windows coated with Al$_2$O$_3$ layers of different thickness (5, 10 and \SI{40}{\nano\meter} corresponding to cells B2, C3 and E3, respectively). Since the initial pressure was higher, the trace regarding cells fabricated in the framework of Ref.~\cite{Carle:JAP:2023}, i.e. the uncoated cells and cells coated with \SI{20}{\nano\meter} of Al$_2$O$_3$, are derived from the mean permeation rate and an initial pressure set at a reduced value of 45~Torr. The inset provides a more detailed view of the traces obtained for thicker layers. The right vertical axis converts the clock frequency evolution into the corresponding buffer gas pressure decrease. For clarity and easier comparison between all the cells, the initial pressure values were offset to 0 at $t$~=~0~s. Light-colored zones indicate the size of error bars attributed to the buffer gas pressure estimation.}
\label{fig:1}
\end{figure*}

For the study of the influence of the thickness of Al$_2$O$_3$ coatings onto He permeation through the glass of \ac{MEMS} vapor cells, 10 additional Cs-He cells to the ones fabricated in the framework of Ref.~\cite{Carle:JAP:2023},  and extracted from 3 wafers that can gather about 200~cells, were tested. The cell technology is comparable to the one described in Refs.~\cite{Hasegawa:SA:2011, Vicarini:SA:2018, Carle:JAP:2023}. A cell consists of two deep-reactive ion-etching (DRIE)-etched silicon cavities, one being employed to host a pill dispenser used as the alkali source, and laser-activated once the cell is sealed. The cavities are sandwiched between two anodically bonded glass windows to seal the science cavity, which has a diameter of 2000~$\pm$~20~$\mu$m and a length of 1500~$\pm$~10~$\mu$m. The squared dispenser cavity has a cross-section surface of \SI{2.56}{\milli\meter}$^2$. The two glass substrates are made of \ac{BSG} (Borofloat®33 from SCHOTT) and have a thickness of 510~$\pm$~10~$\mu$m. Al$_2$O$_3$ layers used in our experiments were deposited using \ac{ALD} with a commercial Beneq TFS-200 system. The Al$_2$O$_3$ films have been coated on one full side of each glass wafer prior to bonding with silicon. Hence, only the inner side of the glass windows is covered. This slightly differs from~\cite{Woetzel:2013,Karlen:OE:2017} where the deposition intervened after the first bonding onto the silicon/glass preform, and thus led to perform the last bonding with coatings on both preform and cover. In our case, the thickness can then be increased without preventing the bonding step.

The experimental setup and methodology used to characterize He permeation are similar to those described in~\cite{Carle:JAP:2023}. A \ac{CPT} atomic clock platform carries six physics packages, each embedding a microfabricated cell to be sequentially tested in clock operation. Each package is magnetically-shielded by a mu-metal layer and allows the application of a static magnetic field used to isolate the 0-0 clock transition. All the cells are temperature-stabilized at 70$^{\circ}$C. Atoms in the cells are probed by a dual-frequency laser field obtained by modulation at 4.596 GHz of a vertical-cavity surface emitting laser (VCSEL) tuned on the Cs D$_1$ line~\cite{Kroemer:AO:2016}. The \ac{CPT} resonance detected at the cell output by a photodiode is acquired and processed by electronics to operate clock operation. The microwave signal is provided by a commercial frequency synthesizer piloted by an active hydrogen maser, used as a reference for frequency shift measurements. The clock frequency for each cell was monitored over weeks. Helium leaking out of the cell by permeation through the glass windows induces a progressive change of the clock frequency because of the buffer gas pressure-induced collisional shift~\cite{Kozlova:PRA:2011}. From the time constant $\tau$ associated to the exponential-decay of the clock frequency, signature of the gas permeation process~\cite{Abdullah:APL:2015, Dellis:OL:2016, Carle:JAP:2023} in the form of the permeation rate $K$, in m$^2$.s$^{-1}$.Pa$^{-1}$, can be extracted.

\begin{table*}[t!]
\vspace*{-0.2ex} \hspace{0.1ex}
\renewcommand{\arraystretch}{1.5}
\begin{ruledtabular}
\begin{tabular}{c c c c c}
Cell id & Al$_2$O$_3$ thickness (nm) & $\tau$ (days) & $P_0$ (Torr) & $K$ (m$^2$.s$^{-1}$.Pa$^{-1}$)  \\
\hline

A1 & \multirow{3}{*}{0 nm} & 62.4 $\pm$ 9.9  & 64.5 $\pm$ 3.0 & (7.0 $\pm$ 1.3) $\times$ $10^{-19}$ \\
A2 &  & 65.6 $\pm$ 10.1  & 64.5 $\pm$ 3.0 & (6.7 $\pm$ 1.2) $\times$ $10^{-19}$ \\
A3 &  & 62.9 $\pm$ 6.0  & 64.5 $\pm$ 3.0 & (7.1$\pm$ 0.9) $\times$ $10^{-19}$ \\
\hline

B1 & \multirow{3}{*}{5 nm} & 93.8 $\pm$ 9.5  & 44.4 $\pm$ 2.1 & (4.7 $\pm$ 0.6) $\times$ $10^{-19}$ \\
B2 &  & 98.2 $\pm$ 11.3  & 35.8 $\pm$ 1.7 & (4.5 $\pm$ 0.6) $\times$ $10^{-19}$ \\
B3 &  & 109.9 $\pm$ 11.1  & 39.4 $\pm$ 1.9 & (4.0$\pm$ 0.5) $\times$ $10^{-19}$ \\
\hline

C1 & \multirow{3}{*}{10 nm} & 191.1 $\pm$ 19.4  & 47.5 $\pm$ 2.2 & (2.3$\pm$ 0.3) $\times$ $10^{-19}$ \\
C2 &  & 150.4 $\pm$ 14.6  & 47.2 $\pm$ 2.2 & (2.9 $\pm$ 0.4) $\times$ $10^{-19}$ \\
C3 &  & 198.6 $\pm$ 20.6 & 37.3 $\pm$ 1.8 & (2.2 $\pm$ 0.3) $\times$ $10^{-19}$ \\
C4 &  & 192.6 $\pm$ 18.6  & 44.6 $\pm$ 2.1 & (2.3 $\pm$ 0.3) $\times$ $10^{-19}$ \\
\hline

D1 & \multirow{3}{*}{20 nm} & 8480 $\pm$ 775  & 65 $\pm$ 3 & (5.1$\pm$ 0.6) $\times$ $10^{-21}$ \\
D2 &  &  6770 $\pm$ 340  & 64.3 $\pm$ 0.3 & (6.4$\pm$ 0.5) $\times$ $10^{-21}$ \\
D3 &  & 7575 $\pm$ 380 & 64.3 $\pm$ 0.3 & (5.8 $\pm$ 0.4) $\times$ $10^{-21}$ \\
D4 &  & 9900 $\pm$ 910  & 44.6 $\pm$ 2.1 & (4.4 $\pm$ 0.5) $\times$ $10^{-21}$ \\
\hline

E1 & \multirow{3}{*}{40 nm} & 11~400 $\pm$ 1050  & 50.9 $\pm$ 2.4 & (3.8 $\pm$ 0.5) $\times$ $10^{-21}$ \\
E2 &  & 13~400 $\pm$ 1200 & 50.8 $\pm$ 2.4 & (3.3 $\pm$ 0.4) $\times$ $10^{-21}$ \\
E3 &  & 21~200 $\pm$ 1~900 & 50.9 $\pm$ 2.4 & (2.0 $\pm$ 0.2) $\times$ $10^{-21}$ \\

\end{tabular}
\end{ruledtabular}
\caption{Compilation of permeation results on all the tested cells. Respective columns show the cell id, the Al$_2$O$_3$ thickness, the time constant $\tau$ of the permeation process, the initial He pressure $P_0$, and the permeation rate $K$ at 70$^{\circ}$C.}
\label{table:PermTable}
\end{table*}

\begin{figure}[t!]
\centering
\includegraphics[width=\linewidth]{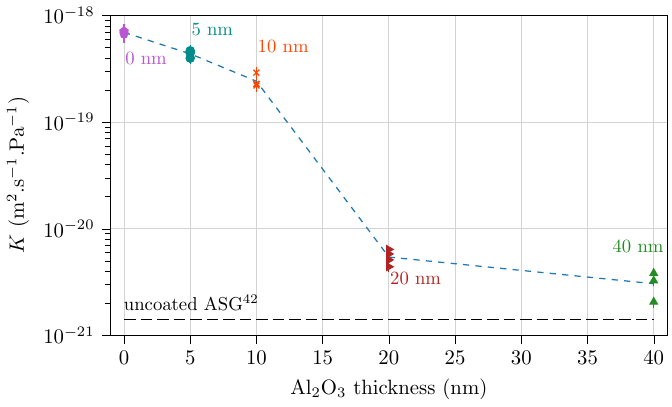}
\caption{Values of the permeation rate $K$ for all tested cells, made of \ac{BSG} windows, versus the Al$_2$O$_3$ layer thickness. The dashed horizontal line indicates the permeation rate obtained with \ac{ASG} substrates, without Al$_2$O$_3$ coatings, reported in \cite{Carle:JAP:2023}.}
\label{fig:2bis}
\end{figure}

Figure~\ref{fig:1} reports the temporal trace of clock frequency measurements performed on one representative cell from each wafer, i.e., one cell for each different thickness of Al$_2$O$_3$ layer (5, 10 and \SI{40}{\nano\meter}). In addition, curves corresponding to an uncoated cell as well as a \SI{20}{\nano\meter}-coated cell are displayed. They are derived from experimental permeation coefficients measured with cells fabricated in the frame of the study reported in Ref.~\cite{Carle:JAP:2023}. Nevertheless, the initial pressure is set at a reduced value of 45~Torr for better comparison of the frequency drift. According to Fig.~\ref{fig:1}, we observe that the shift rate is gradually lowered with increased thickness of the Al$_2$O$_3$ layer for thicknesses up to \SI{20}{\nano\meter}. Interestingly, we do not observe a relevant reduction of the permeation rate for a thickness of \SI{40}{\nano\meter}, in comparison with layers of \SI{20}{\nano\meter}. This suggests that limitation of He permeation requires thicker films of Al$_2$O$_3$ to achieve steady performance compared to addressing the consumption of alkali atoms, for which, at least for Cs, \SI{6}{\nano\meter} appeared sufficient~\cite{Woetzel:2013}. Table~\ref{table:PermTable} summarizes the characteristics of all the tested cells. A slight dispersion is observed for cells coated with the same thickness of Al$_2$O$_3$. The latter might result from a lack of homogeneity of the layers, either during their deposition, or, e.g. during bonding or dispenser activation. 
Figure~\ref{fig:2bis} reports the extracted permeation rate $K$ for all tested cells as a function of the Al$_2$O$_3$ layer thickness. Note that a significant improvement by almost two orders of magnitude is observed between \SI{10}{\nano\meter} and \SI{20}{\nano\meter}. For the best cell coated with \SI{40}{\nano\meter} of Al$_2$O$_3$ (E3), the extracted permeation rate is (2.0~$\pm$ 0.2)~$\times$~10$^{-21}$~m$^2$.s$^{-1}$.Pa$^{-1}$ (at 70$^{\circ}$C). This value is only slightly higher ($\times$~1.4) than the one obtained with uncoated \ac{ASG} substrates~\cite{Carle:JAP:2023}, confirming that Al$_2$O$_3$ layers are efficient alternative gas permeation barriers.
 
\newpage ~
\newpage ~

\section{\label{sec:clock}\ac{CPT} clock with a C\lowercase{s}-N\lowercase{e} microcell, based on A\lowercase{l}$_2$O$_3$-coated \ac{BSG}}
In Ref.~\cite{Carle:OE:2023}, a substantial improvement of the stability of a \ac{CPT} clock was demonstrated, thanks in part to the microcell, fabricated with \ac{ASG} substrates responsible for the reduction of Ne permeation. 

\begin{figure}[h!]
\centering
\includegraphics[width=\linewidth]{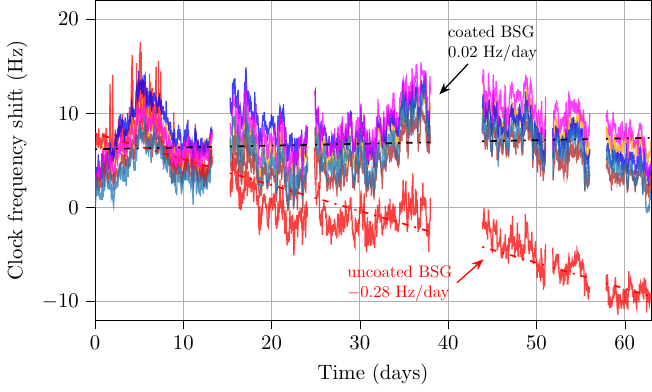}
\caption{Temporal trace of the clock frequency for 5 Cs-Ne microfabricated cells, made of \ac{BSG} wafers coated with \SI{20}{\nano\meter}-thick Al$_2$O$_3$ layers. The trace of the clock frequency obtained for a Cs-Ne cell with uncoated \ac{BSG} windows is shown for comparison. Dashed lines indicate a linear fit to experimental data.}
\label{fig:3}
\end{figure}

Therefore, we have fabricated Cs microcells filled with Ne buffer gas, but based on \ac{BSG} windows coated with \SI{20}{\nano\meter}-thick Al$_2$O$_3$ layers. Five of such cells were tested using the clock setup described in section~\ref{sec:thickness}. Figure~\ref{fig:3} shows the temporal trace of the clock frequency over 63 days. A measurement performed in a Cs-Ne cell built with uncoated \ac{BSG} windows is also reported as a comparison. Despite some perturbations and several interruptions of the measurement (software issues), the long-term frequency variations of the clock appear reduced when the cells are coated with Al$_2$O$_3$ layers. The five tested cells give comparable results, yielding a rate of about 0.02~Hz/day. Conversely, the cell without Al$_2$O$_3$ coating features  a drift rate one order of magnitude higher, of about $-$~0.28~Hz/day. This result is in good agreement with those reported in previous  studies~\cite{Abdullah:APL:2015,Carle:JAP:2023,Carle:OE:2023}, for \ac{BSG} cells filled with neon. 

\begin{figure}[t]
\centering
\includegraphics[width=\linewidth]{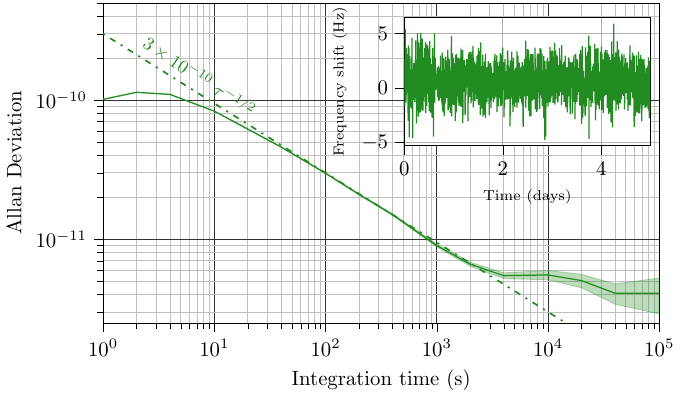}
\caption{Allan deviation of a \ac{CPT} clock based on a Cs-Ne microcell made of \ac{BSG} wafers with 20-nm thick Al$_2$O$_3$ layers. The inset shows the temporal trace of the clock frequency, measured over nearly 5~days, from an initial value of 9192679216 Hz.}
\label{fig:4}
\end{figure}

Furthermore, one of these cells has been tested using the  \ac{CPT} clock setup described in Ref.~\cite{Carle:OE:2023}. The latter employs a pulsed symmetric auto-balanced Ramsey (SABR) interrogation technique~\cite{MAH:APL:2018, MAH:APL:2022} for light-shift mitigation. However, in comparison with studies reported in Ref.~\cite{Carle:OE:2023}, no microwave and laser power servos were applied in the present work. The cell is temperature stabilized at 70$^{\circ}$C. The trace of the clock frequency is shown in the inset of Fig.~\ref{fig:4}, for a duration of about 5~days. The corresponding Allan deviation is reported in Fig.~\ref{fig:4}. The fractional frequency stability is 3~$\times$~10$^{-10}$~$\tau^{-1/2}$ up to a few 1000 s, and reaches a plateau at the level of 4~$\times$~10$^{-12}$ at 4~$\times$~10$^4$~s. According to the study reported in Ref.~\cite{Carle:OE:2023}, this limitation can be attributed to light-shift effects. Despite their presence, the stability at 1~day of 4~$\times$~10$^{-12}$ is significantly lower compared to the typical stability ($\sim$~2-3~$~\times$~10$^{-11}$) achieved with Cs-Ne cells featuring uncoated \ac{BSG} windows~\cite{MAH:APL:2022, Vicarini:UFFC:2019}. 
In other words, if this stability were solely attributable to the permeation of neon, it would correspond to a leak of around $-$0.07~$\mu$bar/day, which is 13 times less than the leak estimated through (uncoated) borosilicate glass by Abdullah~\textit{et al.}~\cite{Abdullah:APL:2015}. It can finally be noted that an additional measurement of the clock frequency with the Al$_2$O$_3$-coated Cs-Ne cell was conducted 6~months later, during 7~days. The same behavior regarding clock stability was observed, validating the long-term effectiveness of Al$_2$O$_3$ coatings. 

\section{\label{sec:conclusions}Conclusions}
This paper reports a study on the influence of the thickness of Al$_2$O$_3$ layers onto the permeation of He gas in microfabricated Cs vapor cells. The cells are based on \ac{BSG} substrates coated with Al$_2$O$_3$ by \ac{ALD}. Permeation rates were derived from measurements of the pressure-shifted transition frequency of a \ac{CPT} atomic clock. A clear reduction of He permeation is observed with increased Al$_2$O$_3$ thicknesses, at least up to \SI{20}{\nano\meter}. Only a slight reduction of He permeation factor is found for thicker layers of \SI{40}{\nano\meter}. In addition, we have tested similar cells, but filled with Ne buffer gas, in 2 different \ac{CPT} atomic clock setups. In particular, we reported the demonstration of a clock with a frequency stability of 4~$\times$~10$^{-12}$ at 10$^5$~s. These performances at 1~day are better than those usually obtained with Cs-Ne microcells made of uncoated \ac{BSG} wafers. These results indicate that Al$_2$O$_3$ layers  offer a promising alternative to \ac{ASG} for reducing gas permeation in \ac{MEMS} cells and are a viable option to enhance the long-term stability of chip-scale atomic clocks and devices.

\section*{Acknowledgments}
This work was supported by the Direction G\'{e}n\'{e}rale de
l’Armement (DGA) and by the Agence Nationale de la Recherche (ANR) in the frame of the ASTRID project named PULSACION under Grant ANR-19-ASTR-0013-01. It was also supported by the Agence Nationale de la Recherche (ANR) in the frame of the LabeX FIRST-TF (Grant ANR 10-LABX-48-01), the EquipX Oscillator-IMP (Grant ANR 11-EQPX-0033) and the EIPHI Graduate school (Grant ANR-17-EURE-0002). The work is also part of the Research Council of Finland Flagship Programme, Photonics Research and Innovation (PREIN), decision number 346545. The PhD thesis of Cl\'{e}ment Carl\'{e} was funded by Centre National d'Etudes Spatiales (CNES) and Agence Innovation D\'{e}fense (AID). Finally, this work was partly supported by the french RENATECH network and its FEMTO-ST technological facility (MIMENTO).

\section*{Author Declarations}
\subsection*{Conflict of Interest}
The authors have no conflicts to disclose.
\subsection*{Author Contributions}
\textbf{Cl\'ement Carl\'e:} Data Curation (lead); Formal Analysis (lead); Methodology (lead); Software (lead); Visualization (lead); Writing/Original Draft Preparation (equal). Writing/Review $\&$ Editing (equal). \textbf{Andrei Mursa:} Resources (equal). \textbf{Petri Karvinen:} Resources (equal), Writing/Review $\&$ Editing (supporting). \textbf{Shervin Keshavarzi:} Resources (equal). \textbf{Moustafa Abdel Hafiz} Formal Analysis (equal); Methodology (equal). \textbf{Vincent Maurice:} Formal Analysis (supporting); Software (equal); Writing/Review $\&$ Editing (supporting). \textbf{Rodolphe Boudot:} Conceptualization (equal); Data Curation (supporting); Formal Analysis (supporting); Funding Acquisition (lead); Methodology (supporting); Project Administration (lead); Visualization (equal); Writing/Original Draft Preparation (lead); Writing/Review $\&$ Editing (supporting). \textbf{Nicolas Passilly:} Conceptualization (lead); Data Curation (equal); Formal Analysis (supporting); Funding Acquisition (equal); Methodology (equal); Project Administration (equal); Resources (lead); Visualization (equal); Writing/Original Draft Preparation (equal); Writing/Review $\&$ Editing (lead).

\section*{Data availability statement}
The data supporting the findings of this study are available from the corresponding author upon reasonable request.

\section*{References}
\bibliography{CARLE_bib}
\end{document}